\pdfoutput=1

\documentclass[11pt,a4paper]{article}

\usepackage{booktabs}

\usepackage[biblatex]{squares}

\bibliography{squares}

\title{Squares: A Fast Counter-Based RNG}

\author{Bernard Widynski}
\mailto{squaresrng@gmail.com}

\begin{document}

\maketitle

\begin{abstract}
In this article, we propose a new counter-based implementation of John von Neumann's middle-square random number generator (RNG).  Several rounds of squaring are applied to a counter to produce a random output.  We discovered that four rounds are sufficient to provide satisfactory data.  Two versions of the RNG are presented, a 4-round version with 32-bit output and a 5-round version with 64-bit output.  Both pass stringent tests of randomness and may be the fastest counter-based generators.
\end{abstract}

\section{Introduction}

In 2011, D. E. Shaw Research published \lq \lq Parallel Random Numbers: As Easy as 1, 2, 3\rq \rq \cite{Salmon}.  A new type of RNG was introduced, the counter-based RNG.  It is distinguished from the conventional RNG in that there is no state.  Random numbers are generated only using a counter.  The Philox \mbox{4x32-10} RNG described in the paper has been installed in \mbox{MATLAB}, NVIDIA's cuRAND, and Intel's MKL.  Philox generates random data with ten rounds of computation.  In this paper, we propose a new RNG that uses John von Neumann's middle-square transformation \cite{Neumann}.  This new RNG which we will call \lq \lq Squares\rq \rq uses only four rounds of squaring.  It is significantly faster than Philox and produces data of equivalent or better quality.

\section{Algorithm}

The squares RNG was derived using ideas from \lq\lq Middle-Square Weyl Sequence RNG\rq\rq \cite{Widynski}.  The msws generator uses a half-square implementation.  That is, only half of the actual square is computed.  The upper bits of this half square are the \lq \lq middle\rq \rq that is returned.  These middle bits are easily obtained by either rotating or shifting the result.  The middle square provides the randomization. Uniformity and period length are obtained by adding in a Weyl sequence.  

For the squares RNG, we replaced the Weyl sequence (w += s) with a counter multiplied by a key.  This turns out to be in effect the same thing. Mathematically, (w += s) is equivalent to \mbox { w = i * s} mod $2^{64}$ for i = 0 to $ 2^{64}-1 $.  That is, i * s will produce the same sequence as (w += s). In place of i and s, we use a counter and a key.   So, if we add counter * key to a square, we should see the same effect as adding a Weyl sequence.  The output will be uniform and $ 2^{64}$  random numbers will be available per key\footnote{The reader is referred to Theorem C in \cite{Widynski}. The probability of any given output is $ 1 / 2^{32} $.  If one generates $2^{64}$ random numbers, one would expect to see $2^{32}$ of each output.}. In the squares RNG, several rounds of squaring and adding are computed and the result is returned. Four  rounds have been shown to be sufficient to pass the statistical tests.  The squares RNG in C is shown below.

\begin{verbatim}

inline static uint32_t squares32(uint64_t ctr, uint64_t key) {
   uint64_t x, y, z;
   y = x = ctr * key; z = y + key;
   x = x*x + y; x = (x>>32) | (x<<32);        /* round 1 */
   x = x*x + z; x = (x>>32) | (x<<32);        /* round 2 */
   x = x*x + y; x = (x>>32) | (x<<32);        /* round 3 */
   return (x*x + z) >> 32;                    /* round 4 */
}
\end{verbatim}

\section{Discussion}

We used the parameters \lq \lq ctr\rq\rq and \lq \lq key\rq\rq to be consistent with Philox parameters.  This generator would be used in a similar way as Philox.

After computing the square, a rotation (circular shift) by 32 bits is performed.  This is done to position the random data into the lower 32 bits which results in a better randomization on the next round.

The key should be an irregular bit pattern with roughly half ones and half zeros.  A utility in the software download\footnote{Software download available at http://squaresrng.wixsite.com/rand}  is provided to create such keys. The keys are chosen so the the upper 8 digits are different and also that the lower 8 digits are different. Different digits assure sufficient change when adding  ctr*key on each invocation of the RNG. The digits are also chosen randomly.  This helps assure that the streams generated will produce relatively random, statistically independent data. The \lq\lq different digits\rq\rq method of initialization was created for the msws RNG in 2017. It has proved to be an effective means of initializing the Weyl sequence.  There have been no reported failures with the msws RNG when using this initialization.  The key utility in the squares RNG software download\footnotemark[2] creates keys using the same different digits method.

For keys generated by the key utility, either ctr*key or (ctr+1)*key will have non-zero digits.  The variables y and z store these values. Adding one or the other will assure non-zero digits in the computation.  This improves the randomization and also provides uniformity.

Since the counter is a 64-bit integer, one can generate $ 2^{64} $ random numbers  per key.  Even on modern super-computers, $ 2^{64} $ is sufficient for most usages.  Assume we have a super-computer with 10 million cores.  If the $ 2^{64} $ is divided equally among all the cores, one could provide a stream of about 1.8 trillion random numbers per core.  Should longer streams be needed, of course, one could use more keys, but it is likely that for most usages a single key would be adequate.  The key utility provided in the software download can provide about 2 billion keys.  This should be a sufficient number for quite some time in the future.

\section{Statistical and Timing Tests}

The test validation for squares was similar to the validation for Philox. D. E. Shaw Research stated that Philox was subjected to at least 89 BigCrush \cite{Lecuyera} tests. For squares, we ran 300 BigCrush tests using random keys.  We also ran inter-stream correlation tests, subset of counter space tests, counter tests with increments other than one, bits-reversed tests, and a basic uniformity test, all with no failures. Additionally, we subjected squares to the \mbox{PractRand} \cite{Dotyhumphrey} test.  We ran 300 PractRand tests with random keys to 256 gigabytes and 25 PractRand tests with random keys to 32 terabytes, all with no failures. 

An additional note on testing:  John Salmon of D. E. Shaw Research suggested that we consider the case where the counter was fixed and only the key changed \cite{Salmon1}.  This is a \lq\lq key counter\rq\rq mode test.  To perform this test we ran the keys utility to generate a keys.h file. The entries from this file were then sent into the generator with the counter fixed. The tests passed even with counter set to zero. The reason is that the digits in keys.h are chosen randomly.  This shows that if keys from the keys utility are used, the streams produced will be statistically independent.  The first outputs from  each stream taken as a set will be random data.

The time to generate one billion random numbers was computed using an Intel Core i7-9700 3.0 GHz processor running Cygwin64 with gcc version 11.2.0.  The time for Philox was 2.29 sec.  The time for squares was 1.35 sec. 

\section{64-Bit Output}

In this section we describe a 64-bit output version of the RNG.  One could of course simply call the 32-bit output version twice to generate a 64-bit output.  This would require 8 rounds of squaring. We discovered that one can produce a satisfactory 64-bit output with just 5 rounds.  This is done by xoring the result of the $4^{th}$ round with a $5^{th}$ round.  See below.

\begin{verbatim}
inline static uint64_t squares64(uint64_t ctr, uint64_t key) {
   uint64_t t, x, y, z;
   y = x = ctr * key; z = y + key;
   x = x*x + y; x = (x>>32) | (x<<32);        /* round 1 */
   x = x*x + z; x = (x>>32) | (x<<32);        /* round 2 */
   x = x*x + y; x = (x>>32) | (x<<32);        /* round 3 */
   t = x = x*x + z; x = (x>>32) | (x<<32);    /* round 4 */
   return t ^ ((x*x + y) >> 32);              /* round 5 */
}
\end{verbatim}

This 64-bit version was subjected to the same tests as the 32-bit version.  25 PractRand tests were run to 32 terabytes with no failures.  The time to generate one billion random numbers was computed using an Intel Core i7-9700 3.0 GHz processor running Cygwin64 with gcc version 11.2.0.  The time for 53-bit precision floating point was 1.93 sec.  If one splits the 64 bits into two 32-bit halves, the time for 32-bit precision floating point was 1.23 sec which is faster than the time of 1.35 sec for the 4-round generator.

\section{Summary}

In this paper, we briefly described a new counter-based RNG called Squares.  Counter-based RNGs have no state and are well suited to parallel computing. Since there are no memory references to a state, counter-based RNGs can be faster than a conventional RNG with a state.  Many of the ideas from the \lq \lq Middle-Square Weyl Sequence RNG\rq\rq were used in this new generator.  We discovered that with only four rounds of squaring we could obtain satisfactory data.   The squares RNG was subjected to similar testing as Philox and shows almost twice the speed with equivalent or better data.\newline\newline
A software package with example programs is available at \newline
http://squaresrng.wixsite.com/rand

\paragraph{Acknowledgements}

I would like to thank D. E. Shaw Research for creating the counter-based RNG.  To those people who helped bring about the Middle-Square Weyl Sequence RNG, I remain thankful.

Also, I think I might mention the following.  I had not actually been working on RNGs for some time.  I was daydreaming and reminiscing about things I had worked on and for some reason remembered Prof. Knuth, the author of \lq \lq The Art of Computer Programming\rq \rq \cite{Knuth}.  I soon found myself questioning if one could create a counter-based RNG using the middle square.  I tried some ideas on my home computer and after a few attempts had passed BigCrush. Merely remembering Prof. Knuth led to this investigation.  We didn't actually interact.  Nevertheless, I think I should give Knuth credit for inspiration.  

I should mention this as well. Similar to the Middle-Square Weyl Sequence RNG, I encountered a problem with uniformity.  One case had too many zeros in the output.  I didn't see an easy fix.  However, I was reminded of the 3n+1 problem, also known as the Ulam conjecture or Collatz conjecture.  After remembering this, a simple solution occurred to me.  All we needed to do was add (ctr+1) * key.  This solved the problem.  The similarity is the +1.   Again, no actual interaction, but I think that Ulam in some indirect sense inspired this solution and I am thankful.

Lastly, I would like to thank Justin Lebar who detected a statistical failure with the 3-round version of the RNG.  There are no known failures with the 4-round or 5-round versions presented in this paper.

\ifx\printbibliography\undefined
    \bibliographystyle{plain}
    \bibliography{squares}
\else\printbibliography\fi

\end{document}